\documentclass[prl,twocolumn,floatfix,showpacs,superscriptaddress]{revtex4}
\usepackage{amsmath}
\usepackage{graphicx}
\begin{document}
\title{Strong electron correlation effects in non-volatile electronic memory devices}
\author{M.~J.~Rozenberg}
\affiliation{CPHT, Ecole Polytechnique, 91128 Palaiseau Cedex, France.}
\affiliation{Departamento de F\'{\i}sica Juan Jos\'e Giambiagi, 
FCEN, Universidad de Buenos Aires,
Ciudad Universitaria Pabell\'on I, (1428) Buenos Aires, Argentina.}
\author{I.~H.~Inoue}
\affiliation{Correlated Electron Research Center (CERC),
National Institute of Advanced Industrial Science
and Technology (AIST), Tsukuba 305-8562, Japan}
\author{M.~J.~S\'anchez}
\affiliation{Centro At\'omico Bariloche, (8400) San Carlos de Bariloche,
Argentina.}
\date{\today}
\begin{abstract}
We investigate hysteresis effects in a model for non-volatile
memory devices. Two mechanisms are found to produce hysteresis
effects qualitatively similar to those often experimentally
observed in heterostructures of transition metal oxides. One of
them is a novel switching effect based on a metal-insulator
transition due to strong electron correlations at the
dielectric/metal interface. The observed resistance switching
phenomenon could be the experimental realisation of a novel type
of strongly correlated electron device.
\end{abstract}
\maketitle
The inclusion of strong correlation effects in semiconductor
electronic devices has long been a goal of solid state physicists.
Significant progress has been achieved in recent years in studies
of metal-insulator-metal (MIM) structures that display different
types of resistance switching by the application of a voltage or
current pulse; a prototype of two-terminal nonvolatile random
access memories. The insulator of the MIM structure is either a
transition metal oxide\cite{blom,beck,watanabe,liu,tulina} or an
organic material\cite{taylor,ma}, in which the electron
correlations usually play an important role. The switching is
directly related to the large hysteresis observed in the {\sl I-V}
characteristics.
These systems are collectively called resistance random access
memories (RRAM) and the most fundamental current issues for this technology 
are the different hysteretic behaviour observed in the current-voltage 
characteristics and their relation to non-volatility.

The goal of the present work is to investigate the qualitatively
different hystereses of $I$-$V$ curves in the MIM structures that
have been experimentally observed so far, and to associate each
hysteresis type to an appropriate physical mechanism. Our starting
point is a recently introduced model for the RRAM device\cite{prl}. 
We shall first study the
basic model
predictions for the $I$-$V$ characteristics, and then show 
how it can be
extended, incorporating the physics of strongly correlated
effects.

The model assumes the existence of an insulating (and inert)
medium with a nonpercolative structure of metallic domains that
might correspond to defects, grains, phase separated regions, and
so on\cite{prl}. Smaller domains are closer to the electrodes and
are called ``top'' and ``bottom''. A large domain occupies the
bulk of the system that is called ``central''. Carriers tunnel
between domains under the action of an external electric field (or
applied voltage). The probability of charge transfer depends on
phenomenological parameters such as the tunneling rates, the
number of states in the domains and their occupation (see Ref.\
\cite{prl} for details). The model is mathematically defined by
the following system of rate equations,
\begin{eqnarray}
\frac{dn^b_i}{dt}\! = \!\! \Gamma^{eb}_i N_e n^e (1-n^b_i) f_{eb}(V)
 -\Gamma^{bc}_i n^b_i N_c(1-n^c) f_{bc}(V) \\ 
\frac{dn^c}{dt} \!\!= \! \!\! \sum_i 
[\Gamma^{bc}_i N_b n^b_i \! (\!1\!-\!n^c\!)f_{bc}(V)
- \Gamma^{ct}_i n^c N_t (\!1\!-\!n^t_i\!)\! f_{ct}(V)]   \\
\frac{dn^t_i}{dt}\! = \!\! \Gamma^{ct}_i N_c n^c (1-n^t_i) f_{ct}(V)
- \Gamma^{te}_i n^t_i N_e (1-n^e) f_{te}(V)
\end{eqnarray}

where $\Gamma^{\alpha\beta}_i$ denote the tunneling rates between the
electrodes and domains, and $\alpha,\beta = e,t,c,b$ denote ``electrode",
``top", ``central" and ``bottom". $N_\alpha$ is the total number of
states in electrode or domain $\alpha$, and $n^\alpha$ is the occupation.
$f_{\alpha\beta}(V)$ describes the dependence of the transfer probabilities
between $\alpha$ and $\beta$ on the given external voltage protocol $V(t)$.
The subindex $i$ labels the domains.

To gain a qualitative understanding of the model, we make some simplifying
assumptions. We ignore disorder effects and take
all ``top" and ``bottom" domains to be identical.
The functions $f_{\alpha\beta}(V)$ are assumed to be independent 
of $\alpha$ and $\beta$.
We also take a unique tunneling rate for the
two interfaces $\Gamma^{et} = \Gamma^{be} \equiv \Gamma^{ext}$
and similarly inside the dielectric
$\Gamma^{tc} = \Gamma^{cb} \equiv \Gamma^{int}$.
Therefore, we can just consider the behaviour of the ``average'' top
and bottom domain occupations, that we denote $n^t$ and $n^b$.
Finally the metallic electrodes are assumed half-filled.

The set of equations simplifies to three coupled nonlinear differential equations,
\begin{eqnarray}
\frac{dn^b}{dt}
& = &[\Gamma^{ext} {N_e \over 2} (1-n^b) - \Gamma^{int} n^b N_c(1-n^c)] f(V)\ \ \ \ \\
\frac{dn^c}{dt}
& = &[\Gamma^{int} N_b n^b (1-n^c)- \Gamma^{int} n^c N_t (1-n^t)] f(V) \ \\
\frac{dn^t}{dt}
& = &[\Gamma^{int} N_c n^c (1-n^t)-
\Gamma^{ext} n^t {N_e \over 2}] f(V)\ \ \ \ \ \ \ \ \ \ \ \ \
\end{eqnarray}
that have to
be solved for the unknown $n^\alpha(t)$ with $\alpha=t,c,b$. The coefficients
$f(V)$ are time dependent through the applied voltage protocol $V(t)$.
There is no general solution for this model, but specific cases can be analysed.
As the central domain is taken to be large compared to top and bottom, we may
assume that its occupation remains approximately constant, {\it i.e.},
$\frac{dn^c}{dt} \approx 0$ and $n^c(t) \approx n^c_{\rm o}$.
Under this  condition, the system of Eqs.\,(4)--(6) can be formally solved
and the result for the time dependent bottom domain occupation is,
\begin{equation}
n^b(t) = e^{-\int_0^t\! P(t') dt'}\! \bigg(\int_0^t\!\! Q(t')
e^{\int_0^{t'}\! P(t'') dt''}\! dt' + n^b(0)\! \bigg)
\label{charging}
\end{equation}
where $P(t)= f(V) [\Gamma^{int} N_c (1-n^c_{\rm o}) + \Gamma^{ext} N_e/2] $ and
$Q(t) = f(V) [\Gamma^{ext} N_e/2] $. The occupation of top is simply given by,
$n^t(t) = 1 - [(1-n^c_{\rm o})/n^c_{\rm o}] n^b(t)$. Taking the time derivative on both
sides we see that $n^c_{\rm o}$ controls the charge accumulation in the
model, that is, $[dn^t\!/dt-dn^b\!/dt]$.

Equation (\ref{charging}) provides further insight into the
non-volatility properties. As the system is strictly always out of
equilibrium under applied voltage, the key quantity to consider is
the relaxation time under either applied bias $V_b$ or during a
voltage pulse $V_p$ ($V_b \ll V_p$). The main dependence of the
relaxation time comes from the first exponential factor of (\ref{charging}). 
With the rather natural assumption that the
tunneling probability dependence is  $f(V) \sim e^V$, one finds that
the ratio of relaxation times is exponential in ($V_p - V_b$).
Thus bias relaxation can be made orders of magnitude slower than
the switching time, which is consistent with the observed behavior
in actual systems. We further note that when the device is
disconnected from the external battery there remains an internal
voltage $V_{int}$ due to the inhomogeneous charge distribution.
The coulombic forces would produce the subsequent relaxation
of this state, and this process is crucial for the long 
time non-volatility characteristics of a system\cite{prl}. 
The previous
analysis implies the existence of a compromise: driving
the system to a strong charge inhomogeneous state (with a strong
or long $V_p$ pulse) would provide a desired large set/reset
resistance ratio but, on the other hand, would also lead to faster
relaxation, {\it i.e.}, shorter non-volatility time. These results
are quite consistent with the investigations of relaxation in
Nb-doped SrTiO$_3$/LaTiO$_{3.5}$/Au heterostructures\cite{mann}.

\begin{figure}
\centerline{\includegraphics[width=7.5cm]{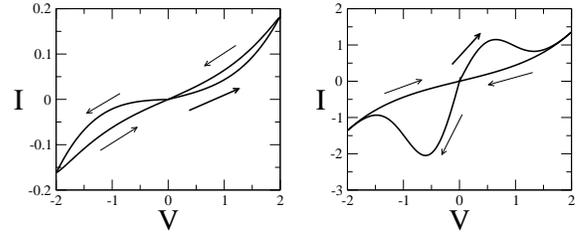}} \caption{
Current-voltage characteristics. $N_t=N_b=10^6$, $N_c=10^{10}$ and
$\Gamma^{int}=2$x$10^{-8}$, $\Gamma^{ext}=6$x$10^{-14}$ (left panel) or
$\Gamma^{int}=8$x$10^{-9}$, $\Gamma^{ext}=6$x$10^{-12}$ (right panel).
The $V$ protocol cycle begins at the thick arrow.}
\label{fig1}
\end{figure}

Let us now turn to the predictions of the model for the $I$-$V$
characteristics and hysteresis effects. The applied voltage
protocol is $0 \rightarrow V_{max} \rightarrow -V_{max}
\rightarrow 0$ with a constant sweep velocity. For simplicity, all
three domains are assumed to be initially half filled. Following
Simmons and Verderber\cite{simmons}, we take the function
$f(V)= \sinh(kV)$ where $k$ depends on various material parameters,
thus we set $k=1$. The (positive) current in and out of the system
is given by the carriers entering the bottom domain and leaving
the top domain respectively. In Fig.\,\ref{fig1} we show the
results for two qualitatively different hystereses that the model
predicts depending on the choice of parameters. Their main
difference is that one shows low to high current switching (left)
while the other shows the opposite behavior (right). It is not
difficult to understand the reason for this contrast. The current
out (for current in, a similar analysis can be done) depends
mainly on the filling level of the top domains. The larger their
occupation, the more carriers that are available to tunnel out.
During the positive side of the voltage protocol, $V$ is initially
increased and then decreased down to zero. During this phase,
charge is transferred from the center domain to the top and from
there to the electrode. Therefore, if more charge enters the top
domain than leaves it, the filling level during the increasing
positive-$V$ ramp would be lower than the filling level during the
subsequent decreasing positive-$V$ ramp. This would lead to
low-to-high current switching as shown in the left panel.
Therefore, the qualitative form of the hysteresis depends on the
intrasystem (top/center and bottom/center) charge transfers
compared to the interface (top/electrode and bottom/electrode)
ones, both of which are controlled by the model parameters
(basically the $\Gamma N$ products). Assuming larger {\it interface}
transfers should be appropriate for systems such as Au/SiO/Au
studied in the pioneering work of Simmons {\it {\sl et
al.}}\cite{simmons,pakhomov}, where hysteresis and negative
differential resistance (NDR) were observed and  qualitative agreement
with our model is found, as shown in the right panel of Fig.\,\ref{fig1}. 
On the other hand, larger {\it intrasystem} transfer is a likely assumption for the
heterostructures with transition metal oxide dielectric, where electronic
inhomogeneous states are often realized. It would also be the case for
the system of artificially created domains of Al-nanoclusters,
where, in fact,   
hysteresis similar to the results shown in the left panel of Fig.\,\ref{fig1}
has been observed\cite{ma}.

While the present model provides useful insight
for the analysis of relaxation times and basic
switching mechanisms\cite{prl}, the qualitative
predictions for hysteresis effect illustrated above show a notable
limitation; namely, the $I$-$V$ curves {\it never cross} at the
origin. Experimentally, {\it crossing} $I$-$V$ characteristics are
often observed\cite{blom,beck,watanabe,tulina,taylor}, thus
providing an important motivation to consider two extensions of
the basic model, Eqs.\,(1)--(3), that we shall describe next.

The first one is to incorporate the physics of interface
tunneling rates that depend on the charge accumulation, {\it
i.e.}, the dependence of $\Gamma^{ext}$ with $n^t$ and $n^b$. 
The underlying physical
mechanism is similar to the dependence of the depletion width with
impurity concentration in ohmic contacts of metal/semiconductor
interfaces\cite{sze}. Basically, the accumulated charge density at
the interface produces additional band bending that leads to a
reduction in tunneling, thus enhancing $\Gamma^{ext}$. Borrowing
from standard junction tunneling theory\cite{sze} we can
qualitatively model this dependence as
$\Gamma^{ext}_{t,b}=e^{-1/\sqrt{n^{t,b}}+1}\Gamma^{ext}$ (Ref.\
\cite{cassey}) so the tunneling is enhanced with increasing
charge accumulation in the small interface domains right up to the
saturation value $\Gamma^{ext}$ for fully occupied domains, {\it
i.e.}, $n^{t,b}$=1. 
This mechanism is widely assumed to be relevant
for many resistance switching phenomenona, however it is also
usually linked to {\it poor non-volatility} characteristics.

\begin{figure}
\centerline{\includegraphics[width=7.5cm]{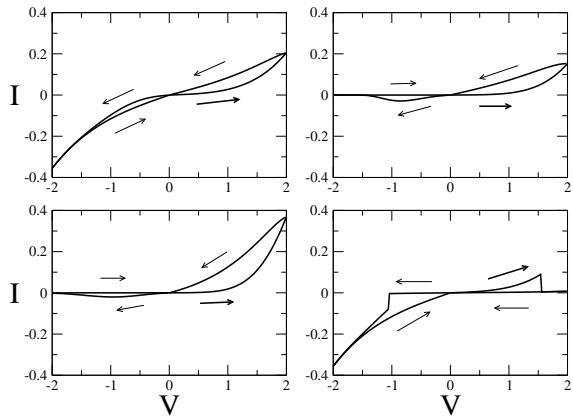}} \caption{
Current-voltage characteristics. $N_t=N_b=10^6$, $N_c=10^{10}$,
$\Gamma^{int}=6$x$10^{-8}$, $\Gamma^{ext}=10^{-13}$ and
$n^t(0)=n^b(0) =n^c(0)=0.1$. Basic model (top left), model with
interface charge dependent tunneling (top left), idem with
Schottky effect (bottom right), model with occupation driven Mott
transition and $\Delta/T =3$ (bottom right). The $V$ protocol
cycle begins at the thick arrow. } \label{fig2}
\end{figure}

We incorporated this effect in our basic set of Eqs.\,(4)--(6)
and solved for the model behavior. The results are shown in Fig.\
\ref{fig2} (top right) along with the predictions of the original
basic model (top left) for the same parameter set (for reference).
We see that, in fact, this additional assumption is sufficient to
turn the hysteresis into the {\it crossing} type, a result that
can be understood through a similar analysis to that we described
earlier\cite{prl}.
An important point to mention here is that the hysteresis effect
does not necessarily need to take place at both electrodes
simultaneously, since the large center domain acts as a buffer
that effectively decouples the two interfaces. Hence this
mechanism would also be relevant for systems with two different
interfaces where the switching properties are controlled by one of
them, {\it e.g.} in heterostructures such as
Au/PbTiO$_3$/La$_{0.5}$Sr$_{0.5}$CoO$_3$ where the Au interface forms a
Schottky barrier but La$_{0.5}$Sr$_{0.5}$CoO$_3$ does not\cite{blom}. 
To investigate a possible effect of Schottky barriers in our results, we modified
the voltage dependent part of the tunneling function to the form
$f(V) = e^V$-1 (forward bias) that produces Schottky type
$I$-$V$ characteristics. The results are shown in Fig.\,\ref{fig2}
(bottom left) and we find no qualitative differences in the
hysteresis.

As we mentioned before, this mechanism is often linked with poor
non-volatile characteristics, possibly because it lacks a
stabilisation effect that could lengthen the lifetimes of the
charge-accumulated (or -depleted) metastable states. A physical
mechanism that would enable such stabilisation is likely to be due
to a {\em coherent} effect in which many particles act in a
correlated manner. Here we propose that such a mechanism is based
on a Mott metal-insulator transition taking place in the domains
when they become close to half filling. In fact, this occupation
(or doping) driven metal-insulator transition is a strong
correlation effect found throughout the transition metal oxide
series\cite{imada} and even in some organic
materials\cite{organic}. For concreteness, we shall assume that
the domains undergo a Mott transition when they are close to
half-filling. On becoming insulating, the domains open an
insulating gap $\Delta$ in their excitation spectrum which we
model by assuming a reduction in the density of states at the
Fermi energy given by $D= e^{-\Delta/T}D_{\rm o}$, $T$ being the
temperature. The dependence of the density of states at the Fermi
energy was entered implicitly in our equations through a
multiplicative factor (set to unity) in the function $f(V)$, since
we were always concerned with tunneling between metallic domains.
We now consider this effect explicitly by modifying $f(V)$ in the
model Eqs.\,(4)--(6) to $f_{mit}(V) = e^{-\Delta/T}f(V)$ if the
occupation of the corresponding domain is within 10\% of
half-filling, or else $f_{mit}(V)=f(V)$. The results are shown in
Fig.\,\ref{fig2} (bottom right). The results can be understood as
follows. Initially, the domains are assumed to be lightly filled,
{\it i.e.} in the metallic state. As the applied voltage is
increased, charge is accumulated at the top domains thus
increasing their occupation level. Eventually, when they approach
half-filling, they become insulating and open a Mott gap. The
tunneling probability to and from the top domain gets
dramatically reduced, with a consequent drop in the output current
and also in the rate of further charge injection. With the reduced
charge injection, the occupation level of the domain remains about
constant and the current is low for the rest of the positive
voltage sweep ($\rightarrow V_{max} \rightarrow 0$). This is also
the case for the beginning of the negative voltage sweep, as the
{\it leakage} of charge out of the top domain is also greatly
suppressed ({\em thus providing an enhanced non-volatility of the
high resistance state}). Eventually a large negative $V$ drives
enough charge out of the top domains and once again they become
metallic, their correlation gap closes and the system switches
back to a high current state.

\begin{figure}
\centerline{\includegraphics[width=7.5cm]{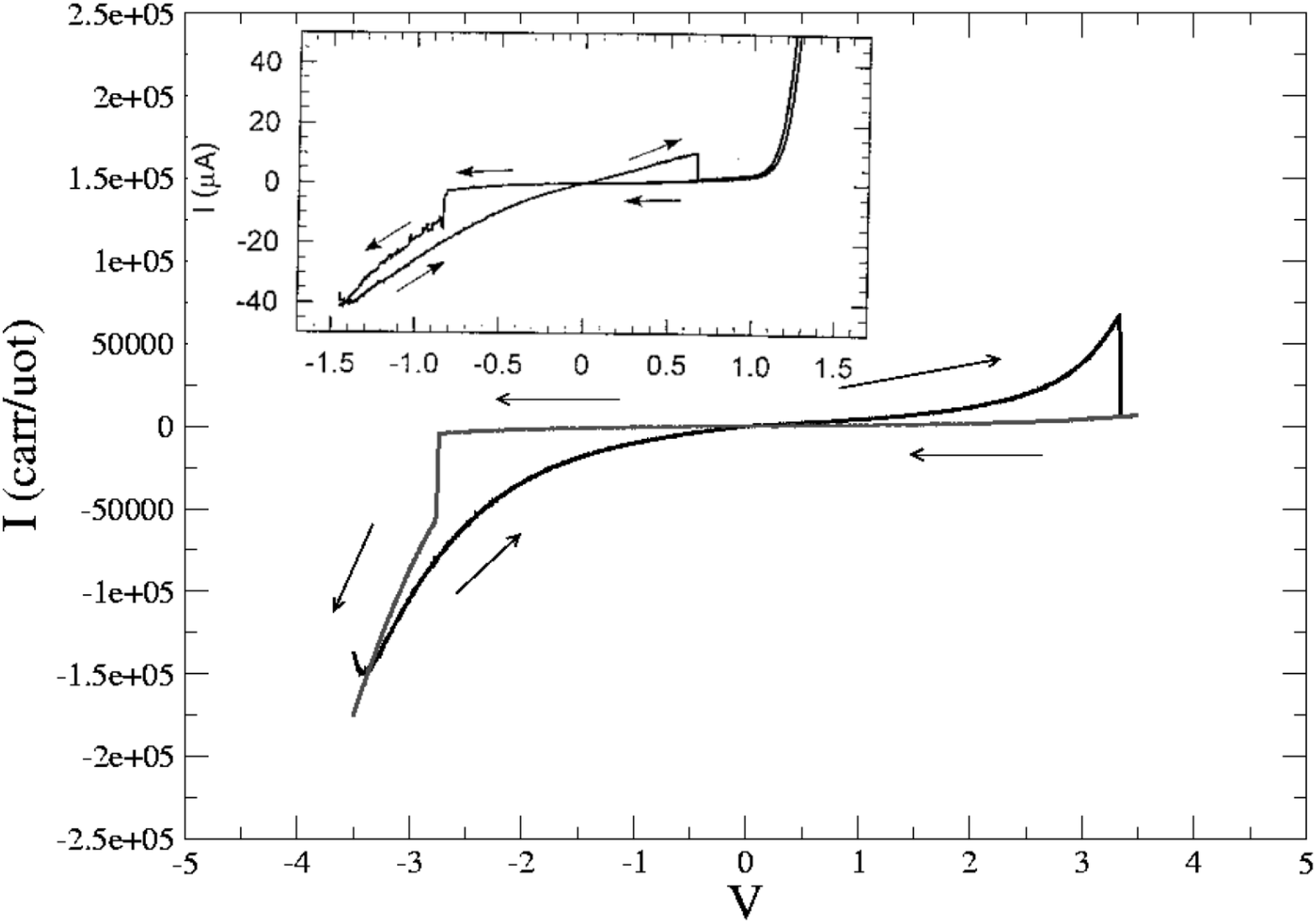}} \caption{
Current-voltage characteristics for the model
with occupation driven Mott transition. The
model parameters are $N_t$=$N_b$=$10^6$, $N_c$=$10^9$ and
$\Gamma^{int}$=$10^{-13}$, $\Gamma^{ext}$=3x10$^{-14}$. 
The current is in carriers per unit of time (ie, Monte
Carlo step).
Inset: experimental $I$-$V$
characteristics in Au/SrTiO$_3$/SrRuO$_3$ from Ref.\
\protect\cite{ibm}. } \label{fig3}
\end{figure}
\begin{figure}
\centerline{\includegraphics[width=7.5cm]{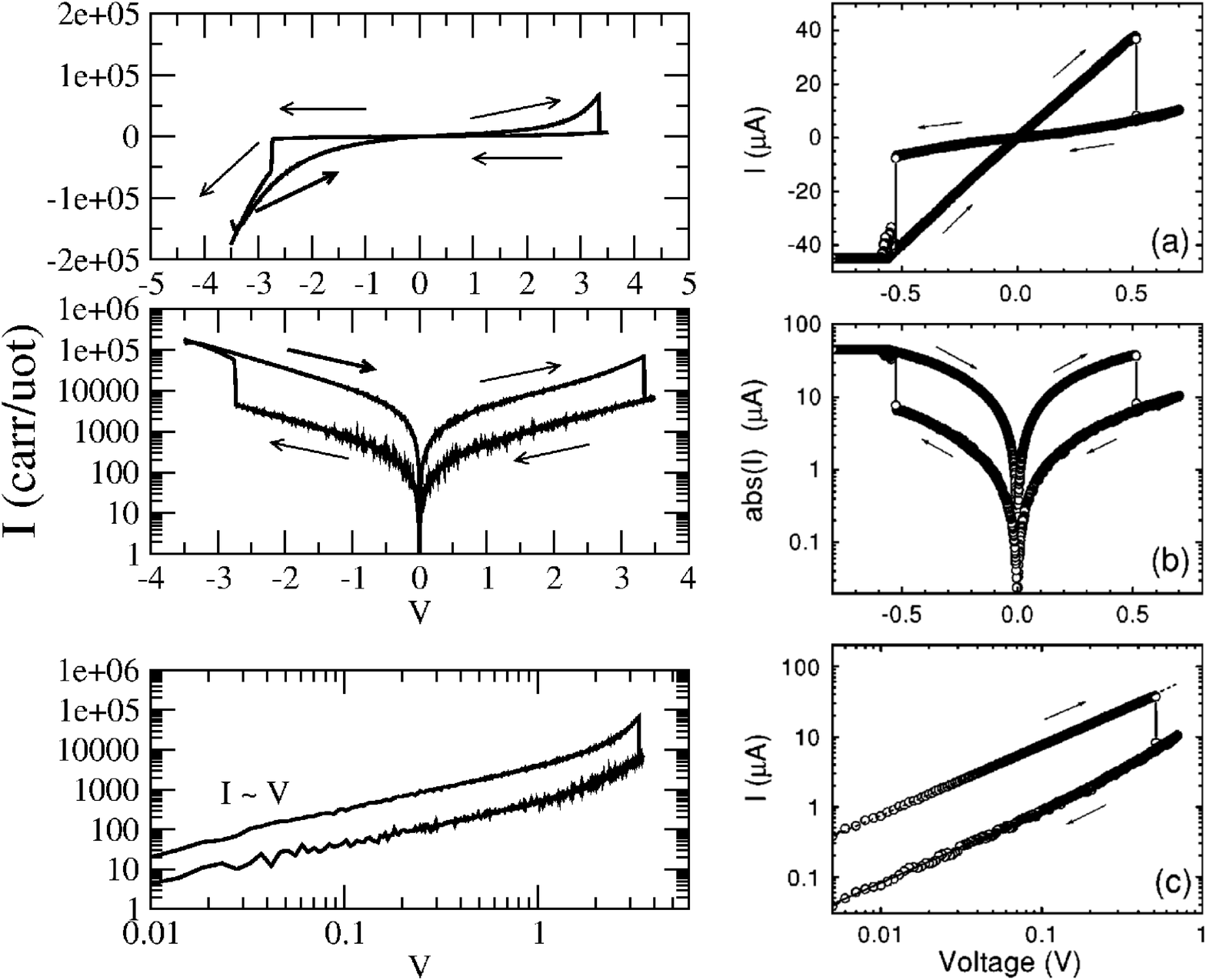}}
\caption{
Idem as in Fig.\,\ref{fig3} in normal scale (top), semi-log scale
(middle) and log-log scale (bottom). The left panels show experiments
in Au/SrZrO$_3$/SrRuO$_3$ from Ref.\ \protect\cite{beck}.
}
\label{fig4}
\end{figure}

In order to give further support to this mechanism we performed
Monte Carlo simulations\cite{prl} on the model given by 
Eqs.\,(1)--(3) supplemented with the Mott transition effect. We looked for
parameters that qualitatively reproduce the experimental results
in the MIM structures with Cr-doped SrTiO$_3$ or SrZrO$_3$ of the
IBM group\cite{beck,ibm}. Our results are shown in Fig.\
\ref{fig3} and the experiments are shown in the inset. (We
modified the voltage protocol to $-V_{max} \rightarrow V_{max}
\rightarrow -V_{max}$ to match the experiment.) Not only is the
qualitative agreement very satisfactory, several details are also
worth pointing out. The initial occupation level of the domains
that allow  best comparison to the experimental data was found to
be small (around 10\%), consistent with the fact that the systems
are nominally empty band insulators with only light doping due to
the Cr substitution. Moreover, the intriguing sudden high to low
current switching at positive applied bias in the experiment is a
feature that naturally emerges from our domain-doping driven metal
to insulator transition scenario. In Fig.\,\ref{fig4} we make more
detailed comparison to the experimental data by plotting the
$I$-$V$ characteristics in different
scale types, all of which show excellent qualitative agreement
with experiment. We note that while the current is proportional to
$V$ at low voltages the system is never truly ohmic in the sense
that there is no actual sample current, rather a series of very
incoherent charge transfer processes. In fact, current conduction
in these systems is known to be non-filamentary\cite{beck}.

To conclude, we have proposed a model for resistance memory
switching and find the different physical mechanisms that lead to
various qualitatively different hysteresis effects in the $I$-$V$
characteristics. We proposed a new scenario for memory switching
in strongly correlated transition metal oxide heterostructures,
based on a metal-insulator transition driven by charge injection
(doping) to domain structures in physical proximity to interfaces.
The information (high resistance state) is stored by driving the
systems to a qualitatively different state which, at the same
time, provides a novel mechanism for the enhancement of the
non-volatility. Our model results are supported by excellent
qualitative agreement with experimental data in various
heterostructures.

We would like to thank Y. Tokura, M. Kawasaki, A. Sawa and A. Odagawa
for valuable comments and discussions. We also acknowledge N. E. Hussey
for critical reading of this paper. Support from Fundaci\'on Antorchas,
UBACyT and ANCTyP is acknowledged.
\end{document}